\documentstyle[12pt]{article}

\newcommand{\be}{\begin{equation}}
\newcommand{\ee}{\end{equation}}

\newcommand{\dlt}{\delta}
\newcommand{\prt}{\partial}
\newcommand{\br}{{\bf r}}
\newcommand{\bk}{{\bf k}}
\newcommand{\bt}{\beta}

\newcommand{\ep}{\varepsilon}
\newcommand{\al}{\alpha}
\newcommand{\ra}{\rightarrow}
\newcommand{\sgm}{\sigma}

\newcommand{\om}{\omega}

\newcommand{\dgr}{\dagger}
\newcommand{\lbd}{\lambda}
\newcommand{\Lbd}{\Lambda}
\newcommand{\cF}{{\cal F}}

\begin{document}

\begin{center}

{\Large {\bf Representative Ensembles in Statistical Mechanics} \\ [5mm]

V.I. Yukalov} \\ [3mm]

{\it Bogolubov Laboratory of Theoretical Physics, \\
Joint Institute for Nuclear Research, Dubna 141980,
Russia}

\end{center}

\begin{abstract}

The notion of representative statistical ensembles, correctly representing
statistical systems, is strictly formulated. This notion allows for a proper
description of statistical systems, avoiding inconsistencies in theory. As
an illustration, a Bose-condensed system is considered. It is shown that a
self-consistent treatment of the latter, using a representative ensemble,
always yields a conserving and gapless theory.

\end{abstract}

\vskip 1cm


\vskip 1cm


\newpage

\section{Introduction}

Statistical systems are characterized by statistical ensembles. It is
crucially important that the given statistical system be correctly represented
by a statistical ensemble. In other words, the chosen statistical ensemble must
be representative for the considered statistical system. This necessitates a
thorough definition of what, actually, a statistical system is, requiring an
accurate enumeration of all its basic features. The usage of a nonrepresentative
ensemble, incorrectly representing the considered statistical system, may lead,
and often does lead, to inconsistencies in the theoretical description of the
system.

The necessity of defining a statistical ensemble that would correctly represent
the given statistical system was, first, emphasized already by Gibbs [1], who
stressed that all additional conditions and constraints, imposed on the system,
must be taken into account. The problem of a proper representation of equilibrium
statistical systems by equilibrium statistical ensembles was discussed by ter
Haar [2,3] and also analized in the review article [4].

The aim of the present paper is to formalize the notion of representative
statistical ensembles by giving precise mathematical definitions and to
generalize this notion for arbitrary systems, whether equilibrium or
nonequilibrium. The application of the notion is illustrated by systems with
Bose-Einstein condensate, when the global gauge symmetry is broken. It is shown
that employing a representative ensemble for a Bose-condensed system results in
the theory enjoying conservation laws and having no gap in the spectrum of
collective excitations.

Throughout the paper, the system of units will be used with the Planck and
Boltzmann constants set to unity, $\hbar=1$, $k_B=1$.

\section{Representative Ensembles}

Let us, first, recall several general preliminary definitions that are necessary
for precisely defining the basic notion of a representative statistical ensemble.

{\it Physical system} is a collection of objects characterized by their typical
features distinguishing this collection from other systems.

For example, a collection of particles can be characterized by their Hamiltonian,
that is, by their energy operator.

{\it Statistical system} is a many-body physical system, whose typical features
are complimented by all additional constraints and conditions which are necessary
for uniquely describing the statistical properties of the system.

Statistical systems are characterized by statistical ensembles.

{\it Statistical ensemble} is a pair $\{ {\cal F},\; \hat\rho(t)\}$ composed by
the space of microstates ${\cal F}$ and a statistical operator $\hat\rho(t)$
on that space.

The space of microstates can be the Fock space or its appropriate subspace.
The statistical operator $\hat\rho(t)$, generally, is a function of time $t$.
To give $\hat\rho(t)$ implies to define its form $\hat\rho(0)$ at the initial
time $t=0$ and to specify the evolution operator $\hat U(t)$ such that
\be
\label{1}
\hat\rho(t) = \hat U(t)\hat\rho(0)\hat U^+(t) \; .
\ee
Therefore, a statistical ensemble can be defined as a triplet
$$
\{ \cF,\hat\rho(0),\hat U(t)\} \; \longleftrightarrow \;
\{ \cF, \hat\rho(t) \} \; .
$$
The knowledge of a statistical ensemble allows one to find statistical
averages.

{\it Statistical average} for an operator $\hat A(t)$ on $\cF$ is
\be
\label{2}
<\hat A(t)> \; = \; {\rm Tr}_\cF \hat\rho(t)\hat A(0) =
{\rm Tr}_\cF \hat\rho(0)\hat A(t) \; .
\ee

Here the Heisenberg representation of the operator $\hat A(t)$ is assumed,
for which
\be
\label{3}
\hat A(t) = \hat U^+(t) \hat A(0) \hat U(t) \; .
\ee

{\it Representative ensemble} is a statistical ensemble equipped with
all additional constraints and conditions that are necessary for a unique
representation of the given statistical system.

Additional constraints and conditions for statistical systems are usually
formulated as conditions on statistical averages for some specified {\it
condition operators} $\hat C_i(t)$, where $i=1,2,\ldots$. These operators do
not need to be necessarily the integrals of motion, but they are supposed to
be Hermitian.

{\it Statistical condition} is a prescribed equality for the statistical
average of a condition operator,
\be
\label{4}
C_i(t) \;  = \; <\hat C_i(t)> \; = \; {\rm Tr}\hat\rho(0)\hat C_i(t) \; .
\ee
Here and in what follows, the trace operation is assumed to be over the
appropriate space of microstates $\cF$.

Let us consider, first, an {\it equilibrium statistical system}, for which
the statistical operator does not depend on time,
\be
\label{5}
\hat\rho(t) = \hat\rho(0) \equiv \hat\rho \; .
\ee
The explicit form of the statistical operator follows from the principle
of minimal information [5]. The latter presumes the conditional maximization
of the Gibbs entropy
\be
\label{6}
S = -{\rm Tr}\hat\rho \ln\hat\rho
\ee
under the statistical conditions (4), among which one usually distinguishes
the definition of the internal energy
\be
\label{7}
E = {\rm Tr}\hat\rho \hat H
\ee
and the normalization condition
\be
\label{8}
{\rm Tr}\hat\rho = 1 \; .
\ee
The {\it information functional} is
\be
\label{9}
I[\hat\rho] = - S +\lbd_0\left ({\rm Tr}\hat\rho -1\right ) +
\bt \left ({\rm Tr}\hat\rho\hat H -  E\right ) + \bt \sum_i \nu_i
\left ( {\rm Tr}\hat\rho \hat C_i - C_i \right ) \; ,
\ee
where $\lbd_0\equiv\ln Z-1$ is the Lagrange multiplier preserving the
normalization condition (8), $\bt$ is the inverse temperature, which is
the Lagrange multiplier for condition (7), and $\bt\nu_i$ are the Lagrange
multipliers related to statistical conditions (4).

The minimization of the information functional (9) yields the statistical
operator
\be
\label{10}
\hat\rho = \frac{1}{Z} \; e^{-\bt H} \; ,
\ee
corresponding to the grand canonical ensemble with the {\it grand Hamiltonian}
\be
\label{11}
H \equiv \hat H + \sum_i \nu_i \hat C_i \; .
\ee
The most customary expression for the grand Hamiltonian (11) is
$$
H = \hat H - \mu\hat N \; ,
$$
where $\mu$ is the chemical potential and $\hat N$ is the number-of-particle
operator. However, the general form of the grand Hamiltonian is given by Eq. 
(11), in which any condition operators can be involved. Thus, an equilibrium
representative ensemble is described by the statistical operator (10) with 
the grand Hamiltonian (11). The evolution operator for an equilibrium system 
is
\be
\label{12}
\hat U(t) = e^{-iHt} \; ,
\ee
which commutes with the statistical operator (10), because of which
$$
i\; \frac{d}{dt} \; \hat\rho(t) =\left [ H,\; \hat\rho(t)\right ] 
= 0 \; .
$$

The general way of obtaining the evolution equations for arbitrary 
nonequilibrium systems is through the extremization of action functionals [6]. 
In our case, this extremization has to be accomplished under the prescribed 
statistical conditions (4).

Let the system Hamiltonian be a functional of the field operators $\psi(x,t)$ 
and $\psi^\dgr(x,t)$, that is, $\hat H=\hat H[\psi]$, where $\psi=\psi(x,t)$. 
The system Lagrangian is
\be
\label{13}
\hat L[\psi] \equiv \int \psi^\dgr(x,t)\; i\; \frac{\prt}{\prt t}\;
\psi(x,t)\; dx - \hat H[\psi] \; .
\ee
The {\it action functional}, or effective action, under the prescribed 
statistical conditions (4), takes the form
\be
\label{14}
A[\psi] \equiv \int \left\{ \hat L[\psi] - \sum_i \nu_i \hat C_i(t)
\right \}\; dt \; ,
\ee
where $\nu_i$ are the Lagrange multipliers guaranteeing the validity of the 
given statistical conditions. The action functional is defined so that to be 
a self-adjoint operator,
\be
\label{15}
A^+[\psi] =  A[\psi] \; .
\ee
Similarly to Eq. (11), the grand Hamiltonian in the Heisenberg representation 
is
\be
\label{16}
H[\psi] = \hat H[\psi] + \sum_i \nu_i \hat C_i(t) \; .
\ee
Then the effective action (14) can be rewritten as
\be
\label{17}
A[\psi] = \int \left\{ \int \psi^\dgr(x,t)\; i\; 
\frac{\prt}{\prt t} \; \psi(x,t)\; dx - H[\psi]\right \} \; dt \; .
\ee
The extremization of the action functional, requiring that
$$
\dlt A[\psi] = 0 \; ,
$$
with
$$
\dlt A[\psi] = \frac{\dlt A[\psi]}{\dlt\psi(x,t)} \;
\dlt\psi(x,t) + \frac{\dlt A[\psi]}{\dlt \psi^\dgr(x,t)}\;
\dlt\psi^\dgr(x,t) \; ,
$$
yields the evolution equations
\be
\label{18}
\frac{\dlt A[\psi]}{\dlt\psi^\dgr(x,t)} = 0 \; , \qquad
\frac{\dlt A[\psi]}{\dlt\psi(x,t)} = 0 \; .
\ee
These equations are the Hermitian conjugated forms of each other.

From Eqs. (17) and (18), it is evident that the evolution equations for the 
field operators can be represented as
\be
\label{19}
i\; \frac{\prt}{\prt t}\; \psi(x,t) = 
\frac{\dlt H[\psi]}{\dlt\psi^\dgr(x,t)}
\ee
and its Hermitian conjugated. This should be equivalent to the Heisenberg 
equation of motion
$$
i\; \frac{\prt}{\prt t}\; \psi(x,t) = 
\left [ \psi(x,t),\; H[\psi]\right ] \; ,
$$
that is, to the Heisenberg representation for the field operator
$$
\psi(x,t) =\hat U^+(t) \psi(x,0) \hat U(t) \; .
$$
Hence, the evolution operator satisfies the Schr\"odinger equation
\be
\label{20}
i\; \frac{d}{d t}\; \hat U(t) =  = H[\psi(x,0)]\; \hat U(t) \; .
\ee

In this way, a nonequilibrium representative ensemble is the set of the 
given space of microstates $\cF$, initial statistical operator $\hat\rho(0)$, 
and of the evolution operator $\hat U(t)$ defined by Eq. (20). An equilibrium 
representative ensemble is, of course, just a particular case of the general 
nonequilibrium ensemble.

\section{Bose-Condensed Systems}

To illustrate the explicit construction of a representative ensemble, let 
us consider a system with Bose-Einstein condensate. Such systems possess 
a variety of interesting properties, as can be inferred from review works 
[7--10].  Moreover, theoretical description of these systems is known to 
confront the notorious difficulty of defining a self-consistent approach. 
The theory of Bose-condensed systems is based on the Bogolubov idea [11-14] 
of breaking the global gauge symmetry by means of the famous Bogolubov 
shift for field operators. The condensate wave function, introduced in 
the course of this shift, has to satisfy the minimum of the related 
thermodynamic potential, which is the stability condition necessary for 
making the system stable and the theory conserving and self-consistent. 
At the same time, the spectrum of collective excitations, according to the 
Hugenholtz-Pines theorem [15], has to be gapless. The notorious problem 
is the appearance of the contradiction between the above two requirements, 
when the theory is either nonconserving or gapful. This contradiction does 
not arise only in the lowest orders with respect to particle interactions, 
when one uses the Bogolubov approximation at low temperatures [11,12] or 
the quasiclassical approximation at high temperatures [16]. However, this 
contradiction immediately arises as soon as the interaction strength is 
not asymptotically weak and one has to invoke a more elaborate 
approximation. This problem of conserving versus gapless approximations 
was first emphasized by Hohenberg and Martin [17] and recently covered 
comprehensively by Andersen [9]. The problem is caused by the usage of 
nonrepresentative ensembles, which renders the system unstable [18]. Here 
we show that employing a representative ensemble never yields the above 
contradiction, always resulting in a self-consistent theory, being both 
conserving and gapless.

We consider a system with the Hamiltonian
$$
\hat H = \int \psi^\dgr(\br) \left ( -\; \frac{\nabla^2}{2m} + U
\right ) \psi(\br)\; d\br +
$$
\be
\label{21}
+ \frac{1}{2} \; \int \psi^\dgr(\br) \psi^\dgr(\br') \Phi(\br-\br') 
\psi(\br')\psi(\br) \; d\br d\br' \; ,
\ee
in which the field operators $\psi(\br)=\psi(\br,t)$ satisfy the Bose
commutation relations, $U=U(\br,t)$ is an external field, and $\Phi(\br)=
\Phi(-\br)$ is an interaction potential. For describing a Bose-condensed 
system with broken global gauge symmetry, the {\it Bogolubov shift} [13,14] 
has to be done through the replacement
\be
\label{22}
\psi(\br,t) \; \longrightarrow \; \hat\psi(\br,t) \equiv 
\eta(\br,t) + \psi_1(\br,t) \; ,
\ee
where $\eta(\br,t)$ is the condensate wave function and $\psi_1(\br,t)$ is 
the field operator of noncondensed particles. The latter field variables 
are assumed to be orthogonal to each other,
\be
\label{23}
\int \eta^*(\br,t) \psi_1(\br,t) \; d\br = 0 \; .
\ee
It is necessary to emphasize that the Bogolubov shift (22) realizes unitary 
nonequivalent operator representations [18,19]. Accomplishing the Bogolubov 
shift (22) in Hamiltonian (21), as well as in all operators of observables, 
we get the algebra of observables defined on the Fock space $\cF(\psi_1)$ 
generated by the field operators $\psi_1^\dgr(\br)$ (see details in Refs. 
[5,18,19]).

The condensate function is normalized to the number of condensed particles
\be
\label{24}
N_0 = \int |\eta(\br,t)|^2 d\br \; .
\ee
The Bogolubov shift (22) is only rational when the number of condensed 
particles (24) is macroscopic, which means that the limit
$$
\lim_{N\ra\infty} \; \frac{N_0}{N} >  0 
$$
is not zero, where $N$ is the total number of particles. The latter is 
given by the average
\be
\label{25}
N \; = \; <\hat N>
\ee
for the number-of-particle operator
\be
\label{26}
\hat N = \int \hat\psi^\dgr(\br) \hat\psi(\br)\; d\br \; ,
\ee
in which the Bogolubov shift (22) is again assumed. The statistical 
averaging in Eq. (25) and everywhere below is over the Fock space 
$\cF(\psi_1)$.

Substituting the Bogolubov shift (22) into Hamiltonian (21) gives in 
the latter the terms linear in $\psi_1$, because of which the average 
$<\psi_1>$ can be nonzero. This, however, would result in the 
nonconservation of quantum numbers, e.g., of spin or momentum. 
Therefore, one has to impose the constraint for the conservation of 
quantum numbers,
\be
\label{27}
<\psi_1(\br,t)>\; = \; 0 \; .
\ee
Defining the self-adjoint condition operator
\be
\label{28}
\hat\Lbd(t) \equiv \int \left [ \lbd(\br,t) \psi_1^\dgr(\br,t) +
\lbd^*(\br,t)\psi_1(\br,t)\right ]\; d\br \; ,
\ee
in which $\lbd(\br,t)$ is a complex function, we may represent 
constraint (27) as the {\it quantum conservation condition}
\be
\label{29}
<\hat\Lbd(t)> \; = \; 0 \; .
\ee

In this way, there are three statistical conditions. The first condition 
is the normalization (24) for the number of condensed particles. 
Condition (24) can be represented in the standard form (4) by defining 
the operator
\be
\label{30}
\hat N_0 \equiv \hat 1 \int |\eta(\br,t)|^2 d\br \; ,
\ee
in which $\hat 1$ is the unity operator in the Fock space $\cF(\psi_1)$.
Then Eq. (24) reduces to the statistical condition
\be
\label{31}
N_0 \; = \; <\hat N_0> \; .
\ee
The second condition is the normalization (25) for the total number of 
particles. Equivalently, instead of normalization (25), we may consider the 
normalization condition
\be
\label{32}
N_1 = \; <\hat N_1> \; , \qquad
\hat N_1 \equiv \int \psi_1^\dgr(\br)\psi_1(\br)\; d\br
\ee
for the number of uncondensed particles $N_1=N-N_0$. And the third 
condition is the conservation condition (29). Respectively, the effective 
action, which is now a functional of the two field variables, $\eta(\br,t)$ 
and $\psi_1(\br,t)$, with taking account of the statistical conditions (29), 
(31), and (32), becomes
\be
\label{33}
A[\eta,\;\psi_1] = \int \left ( \hat L + \mu_0\hat N_0 +
\mu_1\hat N_1 +\hat\Lbd \right )\; dt \; .
\ee
Here $\hat L=\hat L[\hat\psi]$ is the Lagrangian (13) under the Bogolubov 
shift (22) and $\hat\Lbd=\hat\Lbd(t)$ from Eq. (28). The quantities $\mu_0$, 
$\mu_1$, and $\lbd(\br,t)$ are the Lagrange multipliers guaranteeing the 
validity of the corresponding statistical conditions. Introducing the grand 
Hamiltonian
\be
\label{34}
H[\eta,\;\psi_1] \equiv \hat H - \mu_0\hat N_0 - \mu_1\hat N_1 - 
\hat\Lbd \; ,
\ee
in which $\hat H =\hat H[\hat\psi]$, with shift (22), and the effective 
Lagrangian
\be
\label{35}
L[\eta,\; \psi_1] \equiv \int \left [ \eta^*(\br,t) \; 
i\; \frac{\prt}{\prt t}\; \eta(\br,t) + \psi_1^\dgr(\br,t)\; 
i \; \frac{\prt}{\prt t}\; \psi_1(\br,t) \right ]\; d\br - 
H[\eta,\;\psi_1] \; ,
\ee
for the action functional (33), we get
\be
\label{36}
A[\eta,\;\psi_1] = \int L[\eta,\psi_1]\; dt \; .
\ee

The evolution equations follow from the extremization of the action 
functional (36), that is, from the variations
\be
\label{37}
\frac{\dlt A[\eta,\psi_1]}{\dlt\eta^*(\br,t)} = 0
\ee
and
\be
\label{38}
\frac{\dlt A[\eta,\psi_1]}{\dlt\psi^\dgr_1(\br,t)} = 0 \; .
\ee
These equations, as is clear from Eqs. (35) and (36), are equivalent to 
the equations of motion
\be
\label{39}
i\; \frac{\prt}{\prt t}\; \eta(\br,t) = 
\frac{\dlt H[\eta,\psi_1]}{\dlt\eta^*(\br,t)}
\ee
and
\be
\label{40}
i\; \frac{\prt}{\prt t}\; \psi_1(\br,t) = 
\frac{\dlt H[\eta,\psi_1]}{\dlt\psi_1^\dgr(\br,t)} \; .
\ee
One has to substitute here Hamiltonian (34) under the Bogolubov shift (22).
Accomplishing the variation in Eq. (39), we get
$$
i\; \frac{\prt}{\prt t}\; \eta(\br,t) = \left ( -\; \frac{\nabla^2}{2m} +
U - \mu_0 \right ) \eta(\br) +
$$
\be
\label{41}
+ \int \Phi(\br-\br') \left [ |\eta(\br')|^2 \eta(\br) +\hat X(\br,\br')
\right ]\; d\br' \; ,
\ee
where the time dependence in the right-hand side, for short, is not 
explicitly shown, $U = U(\br,t)$, and the notation for the {\it correlation 
operator} 
$$
\hat X(\br,\br') \equiv \psi_1^\dgr(\br')\psi_1(\br')\eta(\br) +
\psi_1^\dgr(\br')\eta(\br')\psi_1(\br) +
$$
\be
\label{42}
+ \eta^*(\br')\psi_1(\br')\psi_1(\br) + 
\psi_1^\dgr(\br')\psi_1(\br')\psi_1(\br)
\ee
is used. The variation in Eq. (40) gives
$$
i\; \frac{\prt}{\prt t}\; \psi_1(\br,t) = 
\left ( -\; \frac{\nabla^2}{2m} + U - \mu_1 \right ) \psi_1(\br) +
$$
\be
\label{43}
+ \int \Phi(\br-\br') \left [ |\eta(\br')|^2\psi_1(\br) +
\eta^*(\br')\eta(\br)\psi_1(\br') + \eta(\br')\eta(\br)\psi_1^\dgr(\br') + 
\hat X(\br,\br') \right ] \; d\br' \; .
\ee

To get an equation for the condensate wave function, we have to take the 
statistical average of Eq. (41). For this purpose, we introduce the {\it 
normal density matrix}
\be
\label{44}
\rho_1(\br,\br') \; \equiv \; <\psi_1^\dgr(\br')\psi_1(\br)> \; ,
\ee
the {\it anomalous density matrix}
\be
\label{45}
\sgm_1(\br,\br') \; \equiv \; <\psi_1(\br')\psi_1(\br)> \; ,
\ee
and their diagonal elements, giving the density of noncondensed particles
\be
\label{46}
\rho_1(\br) \equiv \rho_1(\br,\br) \; = \; <\psi_1^\dgr(\br)\psi_1(\br)>
\ee
and the anomalous average
\be
\label{47}
\sgm_1(\br) \equiv\sgm_1(\br,\br) \; = \; <\psi_1(\br)\psi_1(\br)> \; .
\ee
The quantity $|\sgm_1(\br)|$ can be interpreted as the density of paired 
particles [19]. The total density of particles
\be
\label{48}
\rho(\br) = \rho_0(\br) +\rho_1(\br)
\ee
consists of the condensate density
\be
\label{49}
\rho_0(\br) \equiv |\eta(\br)|^2
\ee
and the density of noncondensed particles (46). Averaging Eq. (41), we find 
the equation for the condensate wave function
$$
i\; \frac{\prt}{\prt t}\; \eta(\br,t) = \left ( -\; \frac{\nabla^2}{2m} +
U -\mu_0\right )\eta(\br) +
$$
\be
\label{50}
+ \int \Phi(\br-\br') \left [ \rho(\br')\eta(\br) + 
\rho_1(\br,\br')\eta(\br') +\sgm_1(\br,\br') \eta^*(\br') + 
<\psi_1^\dgr(\br')\psi_1(\br')\psi_1(\br) > \right ] d\br' \; .
\ee
Equations (43) and (50) are the basic equations of motion for the field 
variables $\eta(\br,t)$ and $\psi_1(\br,t)$. These equations, according to 
Eqs. (39) and (40), are generated by the variation of the grand Hamiltonian 
(34). The latter, in agreement with Eq. (20), also defines the evolution 
operator $\hat U(t)$, which satisfies the Schr\"odinger equation
$$
i\; \frac{d}{dt}\; \hat U(t) =  H[\eta(\br,0),\; \psi_1(\br,0)] \hat U(t)\; .
$$
Thus, the representative ensemble for a Bose-condensed system is the triplet
$$
\{\cF(\psi_1),\hat\rho(0),\hat U(t)\} \; .
$$

It is important to stress that the so defined representative ensemble 
possesses a principal feature making it different from the standardly used 
ensemble having the sole Lagrange multiplier $\mu_0\equiv\mu_1$. But then 
the normalization condition (24) cannot be guaranteed. Then the evolution 
equation for the condensate wave function is not a result of a variational 
procedure. For an equilibrium system, this means that the number of condensed 
particles $N_0$ does not provide the minimum of a thermodynamic potential, 
which implies the system instability. All notorious inconsistencies in theory,
manifesting themselves in the lack od conservation laws or in the appearance 
of an unphysical gap in the spectrum, are caused by the usage of 
nonrepresentative ensembles.

\section{Green Functions}

The equations of motion (43) and (50) allow us to derive the evolutional 
equations for the Green functions. To this end, we shall use the compact 
notation denoting the set $\{\br_j,t_j\}$ by the sole letter $j$, so that 
the dependence of functions on the spatial and temporal variables looks 
like
$$
f(12\ldots n) \equiv f(\br_1,t_1,\br_2,t_2,\ldots,\br_n,t_n) \; .
$$
The product of the differentials $d\br_j dt_j$ will be denoted as $d(j)$, so 
that
$$
d(12\ldots n) \equiv \prod_{j=1}^n d\br_j\; dt_j \; .
$$
We shall employ the Dirac delta function
$$
\dlt(12) \equiv \dlt(\br_1-\br_2)\; \dlt(t_1-t_2) \; .
$$
For the interaction potential, we shall use the retarded form
\be
\label{51}
\Phi(12) \equiv \Phi(\br_1-\br_2) \dlt(t_1-t_2+0) \; .
\ee

The matrix Green function $G(12)=[G_{\al\bt}(12)]$ is a $2\times 2$ matrix, 
with $\al,\bt=1,2$, and with the following elements:
$$
G_{11}(12) \equiv -i <\hat T\psi_1(1)\psi_1^\dgr(2)> \; , \qquad
G_{12}(12) \equiv -i <\hat T\psi_1(1)\psi_1(2)> \; ,
$$
\be
\label{52}
G_{21}(12) \equiv -i <\hat T\psi_1^\dgr(1)\psi_1^\dgr(2)> \; , \qquad
G_{22}(12) \equiv -i <\hat T\psi_1^\dgr(1)\psi_1(2)> \; ,
\ee
where $\hat T$ is the time-ordering operator.

Let us introduce the operator
\be
\label{53}
\hat K_j \equiv -\; \frac{\nabla_j^2}{2m} + U(j) - \mu_1 \; ,
\ee
the condensate effective potential
\be
\label{54}
V(12) \equiv \dlt(12) \int \Phi(13)|\eta(3)|^2 d(3) +
\Phi(12)\eta(1)\eta^*(2) \; ,
\ee
and let us rewrite the correlation operator (42) in the form
\be
\label{55}
\hat X(12) = \psi_1^\dgr(2)\psi_1(2)\eta(1) + 
\psi_1^\dgr(2)\eta(2)\psi_1(1) + \eta^*(2)\psi_1(2)\psi_1(1) + 
\psi_1^\dgr(2)\psi_1(2)\psi_1(1) \; .
\ee

We also define the matrix correlation function $X(123)=[X_{\al\bt}(123)]$
with the elements:
$$
X_{11}(123) \equiv - <\hat T\hat X(12)\psi_1^\dgr(3)> \; , \qquad
X_{12}(123) \equiv - <\hat T\hat X(12)\psi_1(3)> \; ,
$$
\be
\label{56}
X_{21}(123) \equiv - <\hat T\hat X^+(12)\psi_1^\dgr(3)> \; , \qquad
X_{22}(123) \equiv - <\hat T\hat X^+(12)\psi_1(3)> \; .
\ee
From the equations of motion (43) and (50), we find the equations
$$
\left ( i\; \frac{\prt}{\prt t_1}\; - \; \hat K_1\right ) G_{11}(12) -
\int V(13)G_{11}(32)\; d(3) - 
$$
$$
- \int \Phi(13)\left [ \eta(1)\eta(3)G_{21}(32) + 
i X_{11}(132)\right ]\; d(3) = \dlt(12) \; ,
$$
$$
\left ( i\; \frac{\prt}{\prt t_1}\; - \; \hat K_1\right ) G_{12}(12) -
\int V(13)G_{12}(32)\; d(3) - 
$$
$$
- \int \Phi(13)\left [ \eta(1)\eta(3)G_{22}(32) + 
i X_{12}(132)\right ]\; d(3) = 0 \; ,
$$
$$
\left ( -i\; \frac{\prt}{\prt t_1}\; - \; \hat K_1\right ) G_{21}(12) -
\int V^*(13)G_{21}(32)\; d(3) - 
$$
$$
- \int \Phi(13)\left [ \eta^*(1)\eta^*(3)G_{11}(32) + 
i X_{21}(132)\right ]\; d(3) = 0 \; ,
$$
$$
\left ( - i\; \frac{\prt}{\prt t_1}\; - \; \hat K_1\right ) G_{22}(12) -
\int V^*(13)G_{22}(32)\; d(3) - 
$$
\be
\label{57}
- \int \Phi(13)\left [ \eta^*(1)\eta^*(3)G_{12}(32) + 
i X_{22}(132)\right ]\; d(3) = \dlt(12) \; .
\ee

The self-energy $\Sigma(12)=[\Sigma_{\al\bt}(12)]$ is a matrix whose elements 
are defined by the relations
$$
\int \left [ \Sigma_{11}(13) G_{11}(32) +\Sigma_{12}(13)G_{21}(32)
\right ]\; d(3) = 
$$
$$
= \int V(13)G_{11}(32)\; d(3) + \int \Phi(13)\left [ 
\eta(1)\eta(3)G_{21}(32) + iX_{11}(132)\right ] \; d(3) \; ,
$$
$$
\int \left [ \Sigma_{11}(13) G_{12}(32) +\Sigma_{12}(13)G_{22}(32)
\right ]\; d(3) = 
$$
$$
= \int V(13)G_{12}(32)\; d(3) + \int \Phi(13)\left [ 
\eta(1)\eta(3)G_{22}(32) + iX_{12}(132)\right ] \; d(3) \; ,
$$
$$
\int \left [ \Sigma_{21}(13) G_{11}(32) +\Sigma_{22}(13)G_{21}(32)
\right ]\; d(3) = 
$$
$$
= \int V^*(13)G_{21}(32)\; d(3) + \int \Phi(13)\left [ 
\eta^*(1)\eta^*(3)G_{11}(32) + iX_{21}(132)\right ] \; d(3) \; ,
$$
$$
\int \left [ \Sigma_{21}(13) G_{12}(32) +\Sigma_{22}(13)G_{22}(32)
\right ]\; d(3) = 
$$
\be
\label{58}
= \int V^*(13)G_{22}(32)\; d(3) + \int \Phi(13)\left [ 
\eta^*(1)\eta^*(3)G_{12}(32) + iX_{22}(132)\right ] \; d(3) \; .
\ee

Let us introduce the matrix {\it condensate propagator} 
$C(12)=[C_{\al\bt}(12)]$, with the elements
$$
C_{11}(12) \equiv -i\eta(1)\eta^*(2) \; , \qquad 
C_{12}(12) \equiv -i\eta(1)\eta(2) \; ,
$$
\be
\label{59}
C_{21}(12) \equiv -i\eta^*(1)\eta^*(2) \; , \qquad 
C_{22}(12) \equiv -i\eta^*(1)\eta(2) \; ,
\ee
The latter have the properties
$$
C_{11}(21) = C_{22}(12) \; , \qquad C_{11}(11)= C_{22}(11) \; ,
\qquad C_{12}(21)=C_{12}(12)\; ,
$$
\be
\label{60}
C_{21}(21) = C_{21}(12) \; , \qquad C_{11}^*(12)=-C_{22}(12)\; ,
\qquad C_{12}^*(12)= - C_{21}(12) \; .
\ee

The binary Green function is a matrix $B(123) =[B_{\al\bt}(123)]$,
\be
\label{61}
B(123) = C(12)G(23) - i\rho_0(2)G(13) + X(123) \; ,
\ee
whose elements are
$$
B_{11}(123) \equiv C_{11}(12)G_{11}(23) + C_{11}(22)G_{11}(13) + 
C_{12}(12)G_{21}(23) + X_{11}(123) \; ,
$$
$$
B_{12}(123) \equiv C_{11}(12)G_{12}(23) + C_{11}(22)G_{12}(13) + 
C_{12}(12)G_{22}(23) + X_{12}(123) \; ,
$$
$$
B_{21}(123) \equiv C_{22}(12)G_{21}(23) + C_{22}(22)G_{21}(13) + 
C_{21}(12)G_{11}(23) + X_{21}(123) \; ,
$$
\be
\label{62}
B_{22}(123) \equiv C_{22}(12)G_{22}(23) + C_{22}(22)G_{22}(13) + 
C_{21}(12)G_{12}(23) + X_{22}(123) \; ,
\ee
and where $\rho_0(1)\equiv|\eta(1)|^2$.

With Eqs. (59) and (61), relations (58), defining the self-energy, can 
be rewritten in the matrix form
\be
\label{63}
\int \Sigma(13) G(32)\; d(3) = i \int \Phi(13) B(132)\; d(3) \; .
\ee
Then the equations of motion (57) acquire the matrix representation
\be
\label{64}
\left (\hat\tau_3\; i\; \frac{\prt}{\prt t_1}\; - \; \hat K_1
\right ) G(12) - \int \Sigma(13)G(32)\; d(3) =\dlt(12) \; ,
\ee
in which the delta function $\dlt(12)$ in the right-hand side is assumed 
to be factored with the unity matrix $\hat 1=[\dlt_{\al\bt}]$ and
\begin{eqnarray}
\nonumber
\hat\tau_3 \equiv \left [
\begin{array}{cc}
1 & 0 \\
0 & -1 
\end{array} \right ]
\end{eqnarray}
is a Pauli matrix.

Introducing the inverse propagator 
\be
\label{65}
G^{-1}(12) \equiv \left (\hat\tau_3\; i\; \frac{\prt}{\prt t_1}\; - \;
\hat K_1\right )\dlt(12) - \Sigma(12)
\ee
allows us to transform Eq. (64) into
\be
\label{66}
\int G^{-1}(13) G(32)\; d(3) = \dlt(12) \; .
\ee
An equivalent representation, following from Eq. (66), is
\be
\label{67}
\int G(13) G^{-1}(32)\; d(3) = \dlt(12) \; .
\ee
For the self-energy, using Eq. (63), we have
\be
\label{68}
\Sigma(12) = i \int \Phi(13) B(134) G^{-1}(42)\; d(34) \; .
\ee

The equations for the Green functions are to be complimented by the equation 
for the condensate wave function (50), which, introducing one more anomalous 
average
\be
\label{69}
\xi(12) \; \equiv \; <\psi_1^\dgr(2)\psi_1(2)\psi(1)> \; ,
\ee
can be represented as
$$
i\; \frac{\prt}{\prt t_1}\; \eta(1) = \left [ -\; \frac{\nabla_1^2}{2m}
+U(1)-\mu_0\right ]\eta(1) +
$$
\be
\label{70}
+ \int \Phi(12) \left [ \rho(2)\eta(1) +\rho_1(12)\eta(2) +
\sgm_1(12)\eta^*(2) + \xi(12)\right ]\; d(2) \; .
\ee

It is the equations for the Green functions and the equation for the 
condensate function, which become mutually incompatible in the standard 
approach, while employing the representative ensemble renders the theory 
self-consistent in any approximation.

\section{Theory Self-Consistency}

One usually confronts inconsistencies in theory considering a uniform 
equilibrium Bose-condensed system. Then, in any given approximation, one 
gets either a nonconserving theory, that is, an unstable system, or one 
finds an unphysical gap in the spectrum, which, actually, again corresponds 
to an unstable system [9,18]. To analyze this problem, we pass now to the 
case of an equilibrium uniform system, when $U=0$.

Then we use the Fourier transform for the Green function
$$
G(12) = \int G(\bk,\om) e^{i(\bk\cdot\br_{12}-\om t_{12})}
\frac{d\bk d\om}{(2\pi)^4} \; ,
$$
in which 
$$
\br_{12} \equiv \br_1 - \br_2 \; , \qquad t_{12} \equiv t_1 - t_2 \; .
$$
By their definition in Eq. (52), the Green function elements possess the 
properties
\be
\label{71}
G_{11}(21) = G_{22}(12) \; , \qquad G_{12}(21) = G_{12}(12) \; , 
\qquad G_{21}(21) = G_{21}(12) \; .
\ee
Therefore the corresponding Fourier transforms satisfy the relations
$$
G_{11}(-\bk,-\om) = G_{22}(\bk,\om) \; , \qquad 
G_{12}(-\bk,-\om) = G_{12}(\bk,\om) \; , 
$$
\be
\label{72}
G_{21}(-\bk,-\om) = G_{21}(\bk,\om) \; .
\ee
Assuming that the system is isotropic, one has
\be
\label{73}
G_{\al\bt}(-\bk,\om) = G_{\al\bt}(\bk,\om) 
\ee
for all $\al,\; \bt$. Combining Eqs. (72) and (73), we find
$$
G_{11}(\bk,-\om) = G_{22}(\bk,\om) \; , \qquad 
G_{12}(\bk,-\om) = G_{12}(\bk,\om) \; , 
$$
\be
\label{74}
G_{21}(\bk,-\om) = G_{21}(\bk,\om) \; .
\ee
Also, for a uniform equilibrium system, one has [14] the equality
\be
\label{75}
G_{21}(\bk,\om) = G_{12}(\bk,\om) \; .
\ee

Fourier-transforming the self-energy
$$
\Sigma(12) = \int \Sigma(\bk,\om) e^{i(\bk\cdot\br_{12}-\om t_{12})}
\frac{d\bk d\om}{(2\pi)^4}
$$
and, similarly, the inverse propagator (65), we have for the latter
\be
\label{76}
G^{-1}(\bk,\om) =\hat\tau_3\om \; - \; \frac{k^2}{2m} +\mu -
\Sigma(\bk,\om) \; .
\ee
Then Eq. (67) reduces to
\be
\label{77}
G^{-1}(\bk,\om) G(\bk,\om) = \hat 1 \; ,
\ee
where $\hat 1=[\dlt_{\al\bt}]$.

From Eqs. (76) and (77), it follows that $G^{-1}(\bk,\om)$, hense, also 
$\Sigma(\bk,\om)$, have the same symmetry properties as $G(\bk,\om)$. In 
particular,
$$
\Sigma_{\al\bt}(-\bk,\om) = \Sigma_{\al\bt}(\bk,\om) \; , \qquad
\Sigma_{12}(\bk,-\om) = \Sigma_{12}(\bk,\om) \; , \qquad
\Sigma_{21}(\bk,-\om) = \Sigma_{21}(\bk,\om) \; , 
$$
\be
\label{78}
\Sigma_{21}(\bk,\om) = \Sigma_{12}(\bk,\om) \; , \qquad
\Sigma_{11}(\bk,-\om) = \Sigma_{22}(\bk,\om) \; .
\ee

The matrix equation (77), explicitly, is the system of equations
$$
\left ( \om \; - \; \frac{k^2}{2m} + \mu_1 -\Sigma_{11}\right ) G_{11} -
\Sigma_{12}G_{21} = 1 \; , 
$$
$$
\left ( \om \; - \; \frac{k^2}{2m} + \mu_1 -\Sigma_{11}\right ) G_{12} -
\Sigma_{12}G_{22} = 0 \; ,
$$
$$
\left ( - \om \; - \; \frac{k^2}{2m} + \mu_1 -\Sigma_{22}\right ) G_{21} -
\Sigma_{21}G_{11} = 0 \; , 
$$
\be
\label{79}
\left ( - \om \; - \; \frac{k^2}{2m} + \mu_1 -\Sigma_{22}\right ) G_{22} -
\Sigma_{21}G_{12} = 1 \; ,
\ee
where, for short, $G_{\al\bt}=G_{\al\bt}(\bk,\om)$ and $\Sigma_{\al\bt}=
\Sigma_{\al\bt}(\bk,\om)$. The solutions to these equations are
$$
G_{11}(\bk,\om) = 
\frac{\om+k^2/2m+\Sigma_{11}(\bk,\om)-\mu_1}{D(\bk,\om)} \; ,
$$
\be
\label{80}
G_{12}(\bk,\om) = -\; \frac{\Sigma_{12}(\bk,\om)}{D(\bk,\om)} \; ,
\ee
with the denominator
$$
D(\bk,\om) \equiv \left [ \om\; - \; \frac{k^2}{2m} \; - 
\Sigma_{11}(\bk,\om) +\mu_1 \right ] \times
$$
\be
\label{81}
\times \left [ \om + \frac{k^2}{2m} +\Sigma_{22}(\bk,\om) - \mu_1\right ]
+ \Sigma_{12}^2(\bk,\om) \; .
\ee
The solutions for $\Sigma_{21}(\bk,\om)$ and $G_{22}(\bk,\om)$ are defined 
by the symmetry properties (72) to (75).

The excitation spectrum is given by the poles of the Green functions, that 
is, by the zero of denominator (81),
\be
\label{82}
D(\bk,\ep_k) = 0 \; .
\ee
Equation (82) can be represented as
\be
\label{83}
\ep_k = \frac{1}{2}\left [ \Sigma_{11}(\bk,\ep_k) -
\Sigma_{22}(\bk,\ep_k)\right ] \pm 
\sqrt{\om_k^2-\Sigma_{12}^2(\bk,\ep_k)} \; ,
\ee
with the notation
\be
\label{84}
\om_k \equiv \frac{k^2}{2m} + \frac{1}{2} \left [ \Sigma_{11}(\bk,\ep_k)
+ \Sigma_{22}(\bk,\ep_k)\right ] -\mu_1 \; .
\ee

Denominator (81) enjoys the property
$$
D(\bk,-\om) = D(\bk,\om) \; .
$$
Consequently, if $\ep_k$ is a solution of Eq. (82), then $-\ep_k$ is also 
its solution, which is in agreement with the form of Eq. (83).

For an equilibrium uniform system, the Bogolubov shift (22) is equivalent 
to the separation of the zero-momentum term in the expansion of the field 
operator over plane waves. The shift itself has meaning only under the 
normalization condition (24), in which $N_0\sim N$, that is, the 
zero-momentum state is macroscopically occupied. The latter becomes 
possible when the single particle spectrum touches zero. Therefore, the 
necessary condition for the existence of Bose-Einstein condensate is
\be
\label{85}
\lim_{k\ra 0} \ep_k = 0 \; .
\ee
This is to be complimented by the stability condition
\be
\label{86}
{\rm Re}\; \ep_k \geq 0 \; , \qquad {\rm Im}\; \ep_k \leq 0 \; .
\ee
This condition should be kept in mind when choosing the sign plus in front 
of the square root in spectrum (83).

Taking limit (85) for spectrum (83), we notice that, according to properties 
(78),
\be
\label{87}
\Sigma_{11}(\bk,0) =\Sigma_{22}(\bk,0) \; .
\ee
By using perturbation theory for a stable system, one can show [15] that 
in all orders of the theory
\be
\label{88}
\Sigma_{\al\bt}(0,0) \geq 0 \; .
\ee
Then the necessary condition (85) yields the expression for the chemical 
potential
\be
\label{89}
\mu_1 =\Sigma_{11}(0,0) - \Sigma_{12}(0,0) \; ,
\ee
which is the Hugenholtz-Pines relation [15].

On the other hand, we have Eq. (70) for the condensate wave function. For 
an equilibrium uniform system, with no external potential $U$, all densities 
do not depend on the spatial and temporal variables,
\be
\label{90}
\rho_0(\br) = \rho_0 \; , \qquad \rho_1(\br)=\rho_1 \; , \qquad
\sgm_1(\br) =\sgm_1 \; , \qquad \rho(\br) =\rho \; .
\ee
The condensate wave function reduces to the constant
\be
\label{91}
\eta(\br,t) = \eta =\sqrt{\rho_0} \; .
\ee
Then we substitute into Eq. (70) the Fourier transforms for the interaction 
potential
$$
\Phi(\br) = \int \Phi_k e^{i\bk\cdot\br} \frac{d\bk}{(2\pi)^3} \; ,
$$
for the normal density matrix (44),
$$
\rho_1(\br_1,\br_2) = \int n_k e^{i\bk\cdot\br_{12}} 
\frac{d\bk}{(2\pi)^3} \; ,
$$
and for the anomalous density matrix (45),
$$
\sgm_1(\br_1,\br_2) = \int \sgm_k e^{i\bk\cdot\br_{12}} 
\frac{d\bk}{(2\pi)^3} \; .
$$
Similarly, the Fourier transform for the anomalous average (69) is
$$
\xi_1(\br_1,\br_2) = \int \xi_k e^{i\bk\cdot\br_{12}} 
\frac{d\bk}{(2\pi)^3} \; .
$$
As a result, Eq. (70) gives
\be
\label{92}
\mu_0 = \rho \Phi_0 + \int \left ( n_k +\sgm_k +
\frac{\xi_k}{\sqrt{\rho_0}} \right ) \Phi_k \; 
\frac{d\bk}{(2\pi)^3} \; .
\ee

Generally, expressions (92) and (89) do not coincide with each other, 
their difference being 
$$
\mu_0 - \mu_1 = \rho \Phi_0 + \int \left ( n_k +\sgm_k +
\frac{\xi_k}{\sqrt{\rho_0}} \right ) \Phi_k \; 
\frac{d\bk}{(2\pi)^3}\; -
$$
\be
\label{93}
- \Sigma_{11}(0,0) + \Sigma_{12}(0,0) \; .
\ee
This is the general expression for the difference between the Lagrange 
multipliers $\mu_0$ and $\mu_1$ for an arbitrary equilibrium uniform 
Bose-condensed system.

Usually, one does not distinguish between the Lagrange multipliers 
$\mu_0$ and $\mu_1$, which implies setting $\mu_0-\mu_1\ra 0$. However, 
as is evident from Eq. (93), there is no any reason for requiring that 
this quantity be zero. As an illustration, we may resort to the 
Hartree-Fock-Bogolubov approximation, in which $\xi_k=0$ and
$$
\Sigma_{11}(0,0) = (\rho+\rho_0)\Phi_0 + \int n_k \Phi_k \;
\frac{d\bk}{(2\pi)^3} \; , \qquad
\Sigma_{12}(0,0) = \rho_0\Phi_0 + \int \sgm_k \Phi_k \;
\frac{d\bk}{(2\pi)^3} \; .
$$
Relation (89) then yields
\be
\label{94}
\mu_1 = \rho\Phi_0 + \int (n_k -\sgm_k) \Phi_k\;
\frac{d\bk}{(2\pi)^3} \; .
\ee
The difference of the chemical potentials (93) becomes
\be
\label{95}
\mu_0 - \mu_1 = 2 \int \sgm_k \Phi_k \; \frac{d\bk}{(2\pi)^3} \; ,
\ee
which, certainly, is nonzero [20].

In this way, the introduction of the additional Lagrange multiplier 
makes the theory {\it completely self-consistent}. All inconsistencies 
that often arise in other works, such as the appearance of a gap in the 
spectrum, system instability or a distortion of the phase transition 
order, are caused by neglecting the difference between the multiplier
$\mu_1$ and the multiplier $\mu_0$. It is worth emphasizing that the 
introduction of the Lagrange multiplier $\mu_0$ for preserving the 
normalization condition (24), from the mathematical point of view is 
{\it strictly necessary}. In other case, the employed ensemble would not 
be representative, hence, could not correctly describe the Bose-condensed 
system with broken gauge symmetry.

\vskip 5mm

{\bf Acknowledgement}

\vskip 2mm

I am grateful for many useful discussions to M. Girardeau, R. Graham, 
H. Kleinert, and E.P. Yukalova.

\end{document}